\documentclass[english,aps,prl,twocolumn,preprintnumbers,nofootinbib]{revtex4-1}
\usepackage[T1]{fontenc}
\usepackage[latin9]{inputenc}
\usepackage{amsmath}

\makeatletter
 \@ifundefined{textcolor}{}
 {%
   \definecolor{BLACK}{gray}{0}
   \definecolor{WHITE}{gray}{1}
   \definecolor{RED}{rgb}{1,0,0}
   \definecolor{GREEN}{rgb}{0,1,0}
   \definecolor{BLUE}{rgb}{0,0,1}
   \definecolor{CYAN}{cmyk}{1,0,0,0}
   \definecolor{MAGENTA}{cmyk}{0,1,0,0}
   \definecolor{YELLOW}{cmyk}{0,0,1,0}
 }
\makeatother

\begin{document}

\preprint{IPMU13-0126}

\title{Towards consistent extension of quasidilaton massive gravity}

\author{Antonio De Felice}

\affiliation{ThEP\textquoteright{}s CRL, NEP, The Institute for Fundamental Study,
Naresuan University, Phitsanulok 65000, Thailand}

\affiliation{Thailand Center of Excellence in Physics, Ministry of Education,
Bangkok 10400, Thailand}

%\email{antoniod@nu.ac.th}

\author{Shinji Mukohyama}

\affiliation{Kavli Institute for the Physics and Mathematics of the
Universe (WPI),
Todai Institutes for Advanced Study, University of Tokyo, 5-1-5 Kashiwanoha,
Kashiwa, Chiba 277-8583, Japan}

%\email{shinji.mukohyama@ipmu.jp}

\begin{abstract}
 We present the first example of a unitary theory of Lorentz-invariant 
 massive gravity, with all degrees of freedom propagating on a strictly
 homogeneous and isotropic, self-accelerating de Sitter background. The
 theory is a simple extension of the quasidilaton theory, respecting the
 symmetry of the original theory but allowing for a new type of coupling
 between the massive graviton and the quasidilaton scalar. 
\end{abstract}
\maketitle

\textbf{Introduction.} Since the pioneering work of Fierz and Pauli
in 1939~\cite{Fierz:1939ix}, it has been a long-standing question
in theoretical physics whether a graviton can have a non-vanishing
mass. Recently a fully nonlinear theory of massive gravity was found
by de Rham, Gabadadze and Tolley (dRGT)~\cite{deRham:2010ik,deRham:2010kj}
and has provided a positive answer to this fundamental question.

The study of massive gravity is motivated not only by the above mentioned
theoretical question but also by the observed acceleration of cosmic
expansion, one of the greatest mysteries in modern cosmology. There
is a possibility that a finite graviton mass might be the source of
accelerated expansion of the universe. In this respect, it is important
to establish a theoretically consistent and observationally viable
cosmological scenario in massive gravity. However, it was recently
shown that all homogeneous and isotropic cosmological solutions in
the dRGT theory are unstable~\cite{DeFelice:2012mx}.

This no-go result suggests two possible directions: (i) to break either
homogeneity~\cite{D'Amico:2011jj} or
isotropy~\cite{Gumrukcuoglu:2012aa,DeFelice:2013awa} 
of the cosmological background, or (ii) to extend the
theory~\cite{D'Amico:2012zv,Huang:2012pe} (see also
\cite{DeFelice:2013nba} for a non self-accelerating bi-gravity
extension). The purpose of the present paper is to explore the second
possibility and to establish a stable self-accelerating homogeneous and 
isotropic cosmological solution. The hope is that this theory will
provide a theoretically acceptable setup to start studying the
phenomenology of this theory and its potential imprints in the
experimental data.

\textbf{The model.} The \textit{quasidilaton}, denoted hereafter as
$\sigma$, is an additional scalar field in the context of an extended
dRGT massive gravity~\cite{D'Amico:2012zv}, introduced to realize
a new global symmetry 
\begin{equation}
\sigma\to\sigma+\sigma_{0}\,,\qquad\phi^{a}\to e^{-\sigma_{0}/M_{{\rm Pl}}}\,\phi^{a}\,,\label{eqn:symmetry}
\end{equation}
where $\phi^{a}$ ($a=0,\cdots,3$) are four scalar fields called
\textit{St\"uckelberg fields} and $\sigma_{0}$ is an arbitrary
constant. The theory also enjoys the Poincare symmetry in the space of
St\"uckelberg fields
\begin{equation}
\phi^{a}\to\phi^{a}+c^{a}\,,\qquad\phi^{a}\to\Lambda_{b}^{a}\phi^{b}\,,
\end{equation}
so that $\phi^{a}$ enter the action only through the so
called Minkowski \textit{fiducial metric} defined as 
\begin{equation}
f_{\mu\nu}=\eta_{ab}\partial_{\mu}\phi^{a}\partial_{\nu}\phi^{b}\,.
\end{equation}

We extend the quasidilaton theory by adding a new type of coupling
between the massive graviton and the quasidilaton. This is achieved
by replacing $f_{\mu\nu}$ in the action of the original theory with
\begin{equation}
\tilde{f}_{\mu\nu}\equiv f_{\mu\nu}-\frac{\alpha_{\sigma}}{M_{{\rm Pl}}^{2}m_{g}^{2}}e^{-2\sigma/M_{{\rm Pl}}}\partial_{\mu}\sigma\partial_{\nu}\sigma\,,\label{eq:extended-fiducial}
\end{equation}
where $\alpha_{\sigma}$ is a new coupling constant~\footnote{We expect
$\alpha_{\sigma}=O(1)$. In other words, the (technically natural)
suppression scale of the new term is 
$\Lambda_2\sim (M_{\rm Pl}m_g)^{1/2}$ and thus is higher than 
$\Lambda_3\sim (M_{\rm Pl}m_g^2)^{1/3}$. The reason for this is because 
the original quasidilaton (i.e.\ the theory with $\alpha_{\sigma}=0$) in
the $\Lambda_3$ decoupling limit enjoys an enhanced Galileon
symmetry~\cite{D'Amico:2012zv}.} and $m_g$ is the graviton mass
introduced in (\ref{eq:action}) below. Note that the factor
$e^{-2\sigma/M_{{\rm Pl}}}$ in the second term was introduced 
so that $f_{\mu\nu}$ and $\tilde{f}_{\mu\nu}$ share the same scaling
property under (\ref{eqn:symmetry}): 
\begin{equation}
f_{\mu\nu}\to e^{-2\sigma_{0}/M_{{\rm Pl}}}\, f_{\mu\nu}\,,\quad\tilde{f}_{\mu\nu}\to e^{-2\sigma_{0}/M_{{\rm Pl}}}\,\tilde{f}_{\mu\nu}\,.\label{eq:scaling}
\end{equation}
Having defined $\tilde{f}_{\mu\nu}$ in this way, a building block
for the action of extended quasidilaton massive gravity is constructed
as 
\begin{equation}
\mathcal{K}_{\ \nu}^{\mu}=\delta_{\ \nu}^{\mu}-e^{\sigma/M_{{\rm Pl}}}\left(\sqrt{g^{-1}\tilde{f}}\right)_{\ \ \nu}^{\mu}\,,
\end{equation}
where $g^{-1}$ represents the inverse $g^{\mu\nu}$ of the physical
metric $g_{\mu\nu}$. It is easy to see from (\ref{eq:scaling}) that
the tensor $\mathcal{K}_{\ \nu}^{\mu}$ is invariant under (\ref{eqn:symmetry}).
We then build the following terms, which provide a mass to the
graviton. 
\begin{eqnarray}
\mathcal{L}_{2} & \equiv & \frac{1}{2}\,([\mathcal{K}]^{2}-[\mathcal{K}^{2}])\,,\\
\mathcal{L}_{3} & \equiv & \frac{1}{6}\,([\mathcal{K}]^{3}-3[\mathcal{K}][\mathcal{K}^{2}]+2[\mathcal{K}^{3}])\,,\\
\mathcal{L}_{4} & \equiv & \frac{1}{24}\,([\mathcal{K}]^{4}-6[\mathcal{K}]^{2}[\mathcal{K}^{2}]+3[\mathcal{K}^{2}]^{2}\nonumber \\
 &  & \qquad\qquad{}+8[\mathcal{K}][\mathcal{K}^{3}]-6[\mathcal{K}^{4}])\,,
\end{eqnarray}
where square brackets denote a trace.

Note that the dependence of the extended fiducial metric
(\ref{eq:extended-fiducial}) on the time-derivative of the quasidilaton
alters the Hamiltonian structure of the system, and one might worry
about possible reappearance of the Boulware-Deser (BD)
ghost~\cite{Boulware:1973my}. Fortunately, the type of theory considered
in the present paper falls into a wider class of models that was claimed
to be free from the BD ghost~\cite{Gabadadze:2012tr}.

After introducing a canonical kinetic term for the quasidilaton field
$\sigma$, we are ready to write down the full Lagrangian as 
\begin{eqnarray}
S & = & \frac{M_{{\rm Pl}}^{2}}{2}\int d^{4}x\sqrt{-g}\biggl[R-2\Lambda-\frac{\omega}{M_{{\rm Pl}}^{2}}\partial_{\mu}\sigma\partial^{\mu}\sigma\nonumber \\
 &  &
  +2m_{g}^{2}(\mathcal{L}_{2}+\alpha_{3}\mathcal{L}_{3}+\alpha_{4}\mathcal{L}_{4})\biggr].\label{eq:action}
\end{eqnarray}
This action can be further extended, e.g.\ by introducing shift-symmetric
covariant Galileon-type kinetic terms for the quasidilaton field, or/and
by introducing other massive gravity Lagrangians with different values
of $\alpha_{\sigma}$, $\alpha_{3}$, and $\alpha_{4}$. One can also add
an extra term $\xi\sqrt{-\tilde{f}}e^{4\sigma/M_{\rm Pl}}$ invariant
under (\ref{eqn:symmetry}). In the present paper, however, we shall
focus our attention to the simplest extension provided by
(\ref{eq:action}).

In the limit $\alpha_{\sigma}\to 0$, the action (\ref{eq:action})
reduces to the one in the original theory of quasidilaton, but it was
shown in \cite{Gumrukcuoglu:2013nza,D'Amico:2013kya} that the original
theory suffers from ghost instability in the scalar sector. In the
following we shall show that the inclusion of the $\alpha_{\sigma}$ term
can render the extended quasidilaton theory stable.

\textbf{The background.} Let us consider here a flat 
Friedmann-Lema\^\i tre-Robertson-Walker (FLRW) ansatz for the theory
defined in Eq.~(\ref{eq:action}), that is 
\begin{eqnarray}
ds^{2} & = & -N(t)^{2}dt^{2}+a(t)^{2}\delta_{ij}dx^{i}dx^{j}\,,\\
\phi^{0} & = & \phi^{0}(t)\,,\\
\phi^{i} & = & x^{i}\,,\\
\sigma & = & \bar{\sigma}(t)\,.
\end{eqnarray}
The extended fiducial metric (\ref{eq:extended-fiducial}) then reduces
to 
\begin{equation}
\tilde{f}_{00}=-n(t)^2\,, \quad \tilde{f}_{ij}=\delta_{ij}\,,
\end{equation}
where
\begin{equation}
 n(t)^2 \equiv 
\bigl(\dot{\phi}^{0}\bigr)^{2}
 +\frac{\alpha_{\sigma}}{M_{\rm Pl}^2m_g^2}\,e^{-2\bar{\sigma}/M_{\rm Pl}}
 {\dot{\bar{\sigma}}}^2\,.
\label{eq:dphi0}
\end{equation}
We introduce the following quantities characterizing the background
solution. 
\begin{eqnarray}
H & \equiv & \frac{\dot{a}}{Na}\,,\\
X & \equiv & \frac{e^{\bar{\sigma}/M_{\rm Pl}}}{a}\,,\\
r & \equiv & \frac{n}{N}\, a\,.
\end{eqnarray}
We consider here $a$ to be a dimensionless quantity, so as $n$,
$N$, $X$, $\omega$, $r$ and $\alpha_{\sigma}$. Also 
$[\phi^{a}]=M^{-1}$, $[H]=M$, and $[\sigma]=M$. As we shall see below,
the three independent equations of motion for the background allow for
an attractor solution on which $H$, $X$, and $r$ are constants.

Varying the action w.r.t. $\phi^0(t)$ and then setting $n(t)=1$ leads to
\begin{equation}
 \partial_t[ a^4\, X(1-X)J] = 0\,, 
\end{equation}
where
\begin{equation}
 J \equiv 3+3(1-X)\alpha_{3}+(1-X)^{2}\alpha_{4}.
\end{equation}
This implies that $X(1-X)J\propto 1/a^4\to 0$ as the universe expands
(i.e.\ $a\to \infty$). We thus have three cases: $X=0$, $X=1$ and
$J=0$. We would not consider the case with $X=0$ since it would lead to
a strong coupling~\cite{D'Amico:2012zv}. The case with $X=1$ is not
interesting since it does not lead to a self-accelerating solution but
corresponds to a solution driven by the bare cosmological constant 
$\Lambda$. Therefore, in this paper we shall consider the case with 
\begin{equation}
 J = 0. 
\end{equation} 
This, together with
\begin{eqnarray}
r & = & 1+\frac{\omega H^{2}}{m_{g}^{2}X^{2}[\alpha_{3}(X-1)-2]}\,,\\
\left(3-\frac{\omega}{2}\right)H^{2} & = & \Lambda+\Lambda_{X}\,,\label{eq:fried}
\end{eqnarray}
leads to a self-accelerating solution. Here, 
\begin{eqnarray}
\Lambda_{X} & \equiv & m_{g}^{2}(X-1)[6-3X\nonumber \\
 &  & {}+(X-4)(X-1)\alpha_{3}+(X-1)^{2}\alpha_{4}]\,. 
\end{eqnarray}
Eq.~(\ref{eq:fried}), together with the requirement that 
$\partial (H^2)/\partial\Lambda>0$, or, in other words, the positivity
of the effective Newton's constant for the background evolution, implies 
that 
\begin{equation}
\omega<6\,. \label{eqn:omega<6}
\end{equation}

This study shows that it is possible, in general, for this theory
to possess self accelerating solutions with effective cosmological
constant given by $\Lambda_{X}$. It should be noticed that the new
component in the extended fiducial metric (\ref{eq:extended-fiducial}),
i.e.\ the term proportional to $\alpha_{\sigma}$, does not enter in the
background dynamics. However, the parameter $\alpha_{\sigma}$, as we
will see later on, will play a crucial role in order to stabilize the
propagation of the perturbation fields.

\textbf{Scalar perturbations.} We have shown the existence of a
self-accelerating solution for this extended quasidilaton theory. In the
following analysis of perturbations, we choose the unitary gauge: we set
the St\"uckelberg fields to their background values. This choice
completely fixes the gauge freedom. 

As for the scalar sector we introduce the metric in the form
\begin{eqnarray}
\delta g_{00} & = & -2N^{2}\Phi\,,\\
\delta g_{0i} & = & Na\,\partial_{i}B\,,\\
\delta g_{ij} & = & a^{2}\left[2\delta_{ij}\Psi+\left(\partial_{i}\partial_{j}-\frac{1}{3}\,\delta_{ij}\partial_{l}\partial^{l}\right)E\right],
\end{eqnarray}
whereas the quasidilaton field is perturbed as
\begin{equation}
\sigma=\bar{\sigma}+M_{\rm Pl}\,\delta\sigma\,.
\end{equation}
After decomposing each perturbation variable into Fourier modes, and
expanding the action up to second order, we find that $B$ and $\Phi$ do
not have kinetic terms as expected. We can thus integrate them
out. Furthermore, on introducing the field redefinition 
\begin{equation}
\delta\sigma=\Psi+\bar{\delta\sigma}\,,
\end{equation}
we notice that $\Psi$ also becomes an auxiliary field. This feature
is due to the specific structure of the graviton mass term in the
action, which has been constructed in such a way that the Boulware-Deser 
ghost~\cite{Boulware:1973my} is removed. After integrating out the field
$\Psi$ as well, the theory admits only two propagating scalar modes.

\emph{No-ghost condition.} We then obtain the kinetic matrix $K_{IJ}$
($I,J=1,2$) in the total Lagrangian 
$\mathcal{L}\ni K_{11}|\dot{\bar{\delta\sigma}}|^{2}+K_{22}|\dot{E}|^{2}+K_{12}(\dot{\bar{\delta\sigma}}^{\dagger}\dot{E}+\mathrm{h.c.})$. 
In order to avoid a ghost degree of freedom, we demand that 
$\det K_{IJ}>0$ and $K_{22}>0$. By requiring these two inequalities for
all momenta, and noting the condition (\ref{eqn:omega<6}) from the
background evolution, we obtain the following conditions 
\begin{equation}
 0<\omega<6\,, \quad
  X^2 < \frac{\alpha_{\sigma}H^2}{m_g^2} < r^2X^2\,. 
  \label{eq:noghost}
\end{equation}
It should be pointed out that the latter condition implies that $r>1$
(note that $r$ is positive by definition) and that
$\alpha_{\sigma}H^2/m_g^2>0$. In particular, if $\alpha_{\sigma}=0$ 
then there is always a ghost in the scalar 
sector~\cite{Gumrukcuoglu:2013nza,D'Amico:2013kya}. In this sense the
$\alpha_{\sigma}$ term introduced in the present paper plays a key role
to establish the stability of the quasidilaton theory. We also notice
that $(\dot{\phi}^0/n)^2=1-\alpha_{\sigma}H^2/(m_g^2r^2X^2)$ 
and that the last inequality in (\ref{eq:noghost}) is equivalent to
$(\dot{\phi}^2/n)^2>0$. We have thus shown the existence of a parameter
regime in which the scalar sector is free from ghost.

\emph{Speed of propagation.} In order to find the speed of propagation
for the scalar modes, we find it convenient to diagonalize the kinetic
matrix by defining the fields $q_{1,2}$ as
\begin{equation}
\bar{\delta s} \equiv k\, q_{1}\,,\quad
E \equiv\frac{q_{2}}{k^{2}}-\frac{K_{12}}{K_{22}}\, k\, q_{1}\,,
\end{equation}
where $k$ is the size of the comoving momentum. The $k$-dependence in
this field redefinition has been introduced so that, for the new kinetic
matrix, the diagonal elements tend to finite (and non-zero) values for 
large $k$. 

The new kinetic matrix $\mathcal{T}_{IJ}$ is diagonal as
\begin{equation}
\mathcal{L}\ni\mathcal{T}_{11}(t,k)\,|\dot{q}_{1}|^{2}+\mathcal{T}_{22}(t,k)|\dot{q}_{2}|^{2}\,,
\end{equation}
where 
\begin{equation}
\mathcal{T}_{11} = (\det K_{IJ})k^2/K_{22}\,,\quad
\mathcal{T}_{22} = K_{22}/k^4\,,
\end{equation}
and, when the no-ghost conditions (\ref{eq:noghost}) hold we find
\begin{equation}
\mathcal{T}_{11}>0\,,\qquad\mbox{\ensuremath{\mathrm{and}}}\qquad\mathcal{T}_{22}>0\,.
\end{equation}

When $k/a\gg H$ and $k/a\gg m_g$, we can safely ignore time-dependence
of each coefficient in the equations of motion. At the leading order in
large $k$ expansion we thus obtain the following structure of the
equations of motion. 
\begin{eqnarray}
\mathcal{T}_{11}\ddot{q}_{1}+k\mathcal{B}\dot{q}_{2} & \simeq & 0\,,\\
\mathcal{T}_{22}\ddot{q}_{2}-k\mathcal{B}\dot{q}_{1}+k^{2}\mathcal{C}q_{2} & \simeq & 0\,,
\end{eqnarray}
where $\mathcal{T}_{11}$, $\mathcal{T}_{22}$, and other coefficients
$\mathbf{\mathcal{B}}$ and $\mathcal{C}$ are $k$-independent. All other
terms in the equations of motion are suppressed by inverse powers of
$k/(aH)$ or $k/(am_g)$. Then one can read off the speed of propagation
as 
\begin{equation}
c_{s}^{2}=\frac{\mathcal{B}^{2}+\mathcal{C}\,\mathcal{T}_{11}}{\mathcal{T}_{11}\mathcal{T}_{22}}\,\frac{a^{2}}{N^{2}}=1
\end{equation}
for one mode and $c_s^2=0$ for the other mode. Thus, scalar modes with
$k/a\gg \max(H, m_g)$ do not develop gradient instabilities. For a
self-accelerating solution ($\Lambda=0$ and thus $H\sim m_g$) this means
that there is no gradient instability parametrically faster than the
cosmological timescale. Therefore the study of the Laplace instabilities
does not add any new constraint to the model.

\textbf{Vector perturbations.} The vector modes in the theory consist
of the vector modes of the metric tensor, that is
\begin{equation}
\delta g_{0i}=aNB_{i}^{T}\,,\quad\delta g_{ij}=\frac{a^{2}}{2}\,(\partial_{i}E_{j}^{T}+\partial_{j}E_{i}^{T})\,,
\end{equation}
where $\partial^iB^T_i=\partial^iE^T_i=0$. As we have seen, the new
$\alpha_{\sigma}$ term does not affect the background evolution. It does
not affect the vector modes either and the results should agree with the
case with $\alpha_{\sigma}=0$ already studied in
\cite{Gumrukcuoglu:2013nza,D'Amico:2013kya}. In fact we find that the
field $B_{i}$ can be integrated out and the reduced Lagrangian becomes 
\begin{equation}
\mathcal{L}=\frac{M_{\rm Pl}^{2}}{16}\, a^{3}\, N\left[\frac{\mathcal{T}_{V}}{N^{2}}\,|\dot{E}_{i}^{T}|^{2}-k^{2}M_{\mathrm{GW}}^{2}|E_{i}^{T}|^{2}\right],
\end{equation}
where
\begin{eqnarray}
\mathcal{T}_{V} & \equiv & \frac{2k^{2}\omega H^{2}a^{2}}{k^{2}(r^{2}-1)+2\omega H^{2}a^{2}}\,,\\
M_{\mathrm{GW}}^{2} & \equiv & \frac{(r-1)X^{3}m_{g}^{2}}{X-1}+\frac{\omega H^{2}(rX+r-2)}{(X-1)(r-1)}\,.
\end{eqnarray}
The speed of propagation for large $k$ reduces to 
$c_{V}^{2}=(M_{\mathrm{GW}}^{2}/H^{2})\cdot(r^{2}-1)/(2\omega)$. 
Thus the stability for vector modes is ensured if $\mathcal{T}_{V}>0$
and $c_{V}^{2}>0$. These conditions, together with the no-ghost
conditions for the scalar modes (\ref{eq:noghost}), impose 
\begin{equation}
 M_{\mathrm{GW}}^{2}>0\,. \label{eq:MGW2>0}
\end{equation}
This condition does not depend on $\alpha_\sigma$, however it constrain
the other parameters in the theory.

\textbf{Tensor perturbations.} As for the tensor modes, defined in the
metric tensor as 
\begin{equation}
\delta g_{ij}=a^{2}h_{ij}^{TT}\,,
\end{equation}
with $\delta^{ij}h^{TT}_{ij}=0$, and $\partial^j h_{ij}^{TT}=0$,
 we also find the same results as in
\cite{Gumrukcuoglu:2013nza,D'Amico:2013kya}. Namely, their Lagrangian
reduces to
\begin{equation}
\mathcal{L}=\frac{M_{\rm Pl}^{2}}{8}a^{3}N\!\left[\frac{|\dot{h}_{ij}^{TT}|^{2}}{N^{2}}-\left(\frac{k^{2}}{a^{2}}+M_{\mathrm{GW}}^{2}\right)\!|h_{ij}^{TT}|^{2}\right].
\end{equation}
This sector is well behaved and the graviton acquires a mass
$M_{\mathrm{GW}}^{2}$, as expected.

\textbf{Self-acceleration}. For a self-accelerating background without a
bare cosmological constant (i.e.\ setting $\Lambda=0$), all stability
conditions are satisfied if
\begin{eqnarray}
&&\left[0<X<1 \quad\mathrm{and}\quad
 1 < r \leq \bar{r} \quad\mathrm{and}\quad
 0<\omega<6  \right]\nonumber\\
&&\mathrm{or}\quad
\left[0<X<1 \quad\mathrm{and}\quad
 r>\bar{r} \quad\mathrm{and}\quad
 0<\omega<\bar{\omega}  \right]\nonumber\\
&&\mathrm{or}\quad
\left[X>1\quad\mathrm{and}\quad \bar{\omega} < \omega < 6
  \right]\, ,
\end{eqnarray}
where $\bar{r}\equiv\frac{2+X}{1+2X}$ and 
$\bar\omega=\frac{6(r-1)^2X^3}{[r^2+2r-1]X^3-6rX^2+6X+2r-4}$,
provided that $\alpha_{\sigma}$ is chosen to satisfy the second of
(\ref{eq:noghost}).

\textbf{Summary.} We have presented the first example of a unitary
theory of Lorentz-invariant massive gravity, with all degrees of freedom
propagating on a self-accelerating de Sitter background. The theory
is a simple extension of the quasidilaton theory, respecting the symmetry
of the original theory but allowing for a new type of coupling between
the massive graviton and the quasidilaton scalar. We have found that:
(i) there exist non-trivial flat FLRW solutions; (ii) a self-accelerating
de Sitter universe is realized as an attractor of the system; and
(iii) for a range of parameters all degrees of freedom on the attractor
have healthy kinetic terms and there is no gradient instability parametrically
faster than the cosmological time scale.

In \cite{Gumrukcuoglu:2013nza,D'Amico:2013kya} it was shown that 
the self-accelerating solution in the original quasidilaton theory,
even including some additional interactions such as Galileon terms and
Goldstone-type terms, always suffers from ghost instability. Our finding
in the present paper, i.e.\ the stability of the self-accelerating
solution in the extended theory, can be considered as an important step
towards a consistent theory of quasidilaton massive gravity. While it
was argued in \cite{D'Amico:2013kya} that properties of perturbations in 
quasidilaton theories are generically UV sensitive, the existence of a
stable extended theory is quite encouraging, and at the very least
provides an existence proof of an unitary theory with a
self-accelerating cosmological background. The setup in the present
paper also provides a framework in which cosmological and
phenomenological implications of massive gravity can be tested.

\begin{acknowledgments}
 We thank Emir G\"umr\"uk\c{c}\"uo\u{g}lu, Kurt Hinterbichler and Mark 
 Trodden for useful comments. S.M. thanks The Institute for Fundamental
 Study (IF), Naresuan University for hospitality during the beginning of
 this work. The work of S.M. was supported by WPI Initiative, MEXT, 
 Japan. S.M. also acknowledges the support by Grant-in-Aid for
 Scientific Research 24540256 and 21111006. 
\end{acknowledgments}

\end{document}